\documentclass{aa} 
\usepackage{lipsum}
\usepackage{graphicx}
\usepackage{txfonts}
\usepackage{amsmath,amsfonts,amssymb} 
\usepackage{natbib}
\usepackage{multicol,caption}
\usepackage{lipsum}
\usepackage{booktabs}
\usepackage{threeparttable}
\usepackage{multirow}
\usepackage{here}
\usepackage{ftnxtra}
\usepackage{fnpos}
\usepackage{stfloats}

\bibpunct{(}{)}{;}{a}{}{,}
\usepackage[colorlinks=true,urlcolor=blue,linkcolor=blue,citecolor=blue]{hyperref}

\begin{document}

\title{Monitoring of the activity and composition  of comets 41P/Tuttle-Giacobini-Kresak and 45P/Honda-Mrkos-Pajdusakova}

\authorrunning{Moulane et al. 2018}
\titlerunning{TRAPPIST monitorning of comets 41P and 45P}
	
\author{
Y. Moulane\inst{1,2}$^\dagger$, E. Jehin\inst{1}, C. Opitom\inst{3}, F. J. Pozuelos\inst{1}, J. Manfroid\inst{1}, Z. Benkhaldoun\inst{2}, A. Daassou\inst{2} and M. Gillon \inst{1}
	   }
	   
\institute{
Space sciences, Technologies \& Astrophysics Research (STAR) Institute, University of Liège, Liège, Belgium\\
$^\dagger$\email{\color{blue}youssef.moulane@doct.uliege.be}
\and
Oukaimeden Observatory, High Energy Physics and Astrophysics Laboratory, Cadi Ayyad University, Marrakech, Morocco
\and
European Southern Observatory, Alonso de Cordova 3107, Vitacura, Santiago, Chile\\
	     }

\date{Received/accepted}

\abstract{
	
We report on photometry and imaging of the Jupiter Family Comets 41P/Tuttle-Giacobini-Kresak and 45P/Honda-Mrkos-Pajdusakova  with the TRAPPIST-North telescope. We observed 41P on 34 nights from February 16, 2017 to July 27, 2017 pre- and post-perihelion (r$_h$=1.04 au), while we collected data for comet 45P from February 10 to March 30 after perihelion (r$_h$=0.53 au). We computed the production rates of the daughter species OH, NH, CN, C$_3$ and C$_2$ and we measured the dust proxy, Af$\rho$, for both comets.  The peak of water production rate of 41P was  (3.46$\pm$0.20)$\times$10$^{27}$ molecules/s on April 3, 2017 when the comet was at 1.05 au from the Sun. We have shown that the activity of 41P is decreasing by about 30\% to
40\% from one apparition to the next. We measured a mean water production rate for 45P of (1.43$\pm$0.62)$\times$10$^{27}$ molecules/s during a month after perihelion. Our results show that these Jupiter Family Comets had low gas and dust activity and no outburst was detected. Relative abundances, expressed as ratios of production rates and Af$\rho$ parameter with respect to OH and to CN, were compared to those measured in other comets. We found that 41P and 45P have a typical composition in term of carbon bearing species. The study of coma features exhibited by the CN gas species allowed the measurement of the rotation period of 41P, showing a surprisingly large increase of the rotation period from (30$\pm$5) hrs at the end of March to (50$\pm$10) hrs at the end of April, 2017 in agreement with recent observations by other teams.
} 

\keywords{Comets: general - Comets: individual: 41P/Tuttle-Giacobini-Kresak, 45P/Honda-Mrkos-Pajdusakova - Techniques: photometric}

\maketitle

\section{Introduction}

Comets are among the best preserved specimens of the primitive Solar nebula. This status gives them a unique role to understand the origins of the Solar System. The success of the Rosetta mission to the Jupiter Family Comet (hereafter JFC) 67P/Churyumov-Gerasimenko (hereafter 67P) is changing our knowledge of comets \citep{Marty2017}. At the same time it shows that ground based observations are also important, to complete \textit{in situ} data by obtaining information on a larger scale of the coma and for a greater number of comets, which is necessary to extrapolate the results of the entire comet population. The link between the composition of comets and their dynamical history must still be clarified.

41P/Tuttle–Giacobini–Kresák (hereafter 41P) is a JFC  discovered by Horace Parnell Tuttle on May 3, 1858, and re-discovered independently by Michel Giacobini and L’ubor Kresák in 1907 and 1951 respectively. The comet had two close encounters with Jupiter in 1975 and 1988 that have altered its orbit slightly. 
Its perihelion distance is now 1.05 au and its orbital period 5.42 years\footnote{JPL small bodies data browser : \url{https://ssd.jpl.nasa.gov/sbdb.cgi\#top}}.

41P is famous for major outbursts that makes it highly variable in brightness. During its 1973 apparition, 41P briefly reached magnitude 4 after a major outburst \citep{Kresak1974}. The apparition of the comet in late 2000 and early 2001 was not expected to be one of the best, with a magnitude about 12 by early January 2001. During that period, 41P was observed by the Solar and Heliospheric Observatory (SOHO). The water production rate was (5.20$\pm$0.07)$\times$10$^{28}$  molecules/s at perihelion (r$_h$=1.05 au, January 6, 2001). This rate was derived by modeling the observed distribution of atomic hydrogen in the cometary coma \citep{Combi2017}. In its 2006 apparition, the water production rate obtained from SOHO decreased to (1.60$\pm$0.08)$\times$10$^{28}$  molecules/s at perihelion (r$_h$=1.05 au). The last perihelion of the comet was on November 12, 2011 but it was not observed because it was on the far side of the Sun. The heliocentric light curve of \cite{Tancredi2000} indicates that 41P has a small nucleus. \citep{Lamy2004} estimate the nucleus radius to be about 0.7 km, but even this may only be an upper limit. The radius of 41P is less than 70\% of all measured radii of JFCs \citep{Fernandez2013}, so 41P can be considered as a small comet.

45P/Honda-Mrkos-Pajdusakova (hereafter 45P) is a short-period comet that also belongs to the Jupiter Family. It was discovered on December 3, 1948 by Minoru Honda, Antonín Mrkos, and Ludmila Pajdušáková. Since then, it has been observed on every apparition except for its 1959 and 1985 passages, when it was too close to the Sun. Over the last few apparitions, it has been close to Jupiter in 1935 and in 1983. Its perihelion distance is now 0.53 au and its orbital period 5.25 years. 

The recent apparitions of 45P were in 2006 when the comet was on the far side of the Sun around the time of perihelion. The last apparition was in 2011 when it was well positioned for pre-perihelion observations. At this time, the comet approached the Earth within 0.06 au on August 15, 2011. Its perihelion was on September 28, 2011 and the comet reached a peak of magnitude of 7.5. The water production rate obtained from SOHO observatory was around 9$\times$10$^{28}$ molecules/s when the comet was 0.54 au from the Sun and at 0.87 au from the Earth \citep{Combi2017}. For the  nucleus size of 45P, \cite{Lowry2003a} obtained a mean value of r$_n$ = 0.34$\pm$0.01 km from observations performed on two consecutive days during the passage in June, 1999 using the Hubble Space Telescope. The snapshot observation of \cite{Lamy1999} gives a much larger value of r$_n$ = 1.34$\pm$0.55 km, but the large error bar means that r$_n$ could be as small as 0.8 km. Using radar observations, \cite{Lejoly2017} estimated the effective radius of 45P's nucleus to be in the range of 1.2-1.3 km with a rotation period of about 7.5 hrs.
 
In this work, we present the monitoring of the activity and composition for various gas species and the dust of 41P over 5 months and of 45P over 40 days after perihelion with the TRAPPIST-North telescope. In section \ref{sec2}, we will report on our observation strategy and the tools used in the data reduction. In section \ref{sec3}, we followed the evolution of the activity of the different daughter molecules considered in this work, and the dust properties obtained from A($\theta$)f$\rho$ parameter over a wide range of heliocentric distances, and on both sides of perihelion for 41P. Section \ref{sec4} contains an explanation of the comparison of the coma morphology exhibited by various gas species allowing us to determine the rotation period of comet 41P. Finally, in Section \ref{sec5}, we summarize our results for both comets.

\section{Observations and data reduction}
\label{sec2}

TRAPPIST-North ({\bf TRA}nsiting {\bf P}lanets and {\bf P}lanetes{\bf I}mals {\bf S}mall {\bf T}elescope) is a 60-cm robotic telescope that was installed in May, 2016 at the Oukaimeden Observatory at 2750 m of altitude in the Atlas mountains of Morocco \citep{Benkhaldoun2018}. The project is led by the University of Liège (Belgium) and the Cadi Ayad University of Marrakech (Morocco). TRAPPIST-North (TN) is a twin of TRAPPIST-South (TS) which was installed at the ESO La Silla Observatory in 2010 \citep{Jehin2011}. TRAPPIST telescopes are dedicated to the detection and characterization of exoplanets and the study of comets and other small bodies of the Solar System. With TN we can now also observe comets in the Northern hemisphere and have with TS access almost the whole sky regularly. With the two telescopes, we can follow comets over a large part of their orbits. The telescope is equipped with a 2K$\times$2K thermo-electrically cooled Andor IKONL CCD camera with a 20$^\prime \times$20$^\prime$ field of view. We bin the pixels 2 by 2 giving a resulting plate scale of 1.3$^{\prime\prime}$/pixel. We observe bright comets (typically V<12) with HB narrow-band filters isolating the emission bands of OH[310 nm], NH[336 nm], CN[385 nm], C$_3$[405 nm], and C$_2$[515 nm] daughter species as well as emission free dust continuum in three regions covering the optical range BC [445 nm], GC[525 nm], and RC[715 nm]. These narrow-band filters are interference filters, whose design was carefully thought to maximize the fraction of the emission bands encompassed while minimizing the continuum contamination for the gas filters. The continuum filters have to be large enough to provide significant signal but avoid gas contamination at the same time. The final HB filter set was the result of these considerations \citep{Farnham2000}. Images are also taken with broad-band B, V, Rc, and Ic Johnson-Cousin filters \citep{Bessell1990}. All these filters are the same as those used for TRAPPIST-South observations.

We have obtained 600 images of comet 41P with TN over 5 months from February 16, when the comet was at 1.27 au from the Sun and at 0.29 au from the Earth, until July 27, 2017 when the comet was at 1.69 au. Its perihelion was on April 12 at 1.0 au and the comet was at its closest distance to Earth on April 1 at only 0.14 au. We also observed 41P with TS before perihelion on February 25 and on March 8, 2017 with narrow- and broad-filters. In the OH, NH, and RC filters, no signal was detected. Exposure times range from 60 s to 240 s for the broad-band filters, and from 300 s to 1200 s for the narrow-band filters. Most Rc, CN, and C$_2$ images were obtained in sets of 4-12 images per night while most images in other filters were obtained as single images.

We started to collect data for comet 45P  as soon as the comet was visible from Morocco, on February 10. It was about one month after perihelion and at a distance of 0.97 au from the Sun and at its closest distance to Earth, only 0.08 au. With a perihelion at 0.53 au, the comet had a small solar elongation (<30$^\circ$) around perihelion. Most ground-based telescopes are not able to observe this close to the Sun, which leads to a lack of data for this comet at perihelion. During most of our observations, the comet coma was huge and filling most of the field of view as it was very close to Earth. The comet was rising fast and it was soon visible all night. 45P faded rapidly and NH was not detected during our observations. We continued to observe 45P about twice a week until the end of March when the comet was at 1.62 au. 45P was also observed with TS on February 25 and on March 8, 20-21, 2017 with narrow- and broad-filters. No signal was detected in the OH and NH filters. The exposure time ranged from 60 s to 240 s for the broad-band filters, and from 300 s to 1200 s for the narrow-band filters.

Standard procedures were used to calibrate the data by creation of master bias, flat and dark frames. The bias and dark subtraction, as well as the flat-field correction are done using IRAF.  The absolute flux calibration was made using standard stars observed during the same period. The next step was the removal of the sky background coming mainly from atmospheric activity and moon contamination, which are variable and depend on the wavelength. The subtraction of the sky background has to be done with great care for extended objects. We used the procedure developed in previous papers \citep{Opitom2015a,Opitom2015b,Opitom2016}. After the determination of the comet center using the {\it Imcntr} IRAF task, we derived median radial brightness profiles from the gas (and dust) images. Then, we removed the dust contamination from the gas radial profiles using images of comets in the BC filter, because it is less contaminated by cometary gas emission than the other dust filters \citep{Farnham2000}. 

The fluorescence efficiency of each molecule, also called g-factor, is used to convert the flux to column density. The g-factor represents the number of photons per second scattered by a single atom or molecule exposed to the unattenuated sunlight \citep{Krishna2010}.

\begin{equation}
g_\lambda =\lambda^2 f_\lambda F_\lambda \frac{\pi e^2}{m_e c^2}\frac{A_{ik}}{\sum_k A_{ik} }
\end{equation}

Where $F_\lambda$ is the solar flux per unit of wavelength and $f_\lambda$ is the oscillator strength. $A_{ik}$ are the Einstein coefficients. The term $\frac{A_{ik}}{\sum_k A_{ik} }$ represents the fraction of photons emitted from the energy level {\it i} to the energy level {\it k} relatively to all the photons emitted from the energy level {\it i}. $e$ and $m_e$ are the charge and mass of electron, respectively. The fluorescence efficiencies values for the different gas filters are taken from Schleicher’s website\footnote{Comet fluorescence efficiency: \url{http://asteroid.lowell.edu/comet/gfactor.html}}. The C$_2$ g-factors have a single value for the $\Delta\upsilon$=0 band sequence with a band head near 5160 $\AA$, and it is scaled by r$_h^{-2}$ \citep{A'Hearn1984}. The C$_3$ g-factor has a single value for the C$_3$ band complex which peaks near 4030 $\AA$ and extends from 3300 to 4400 $\AA$. The C$_3$ fluorescence efficiencies are scaled by r$_h^{-2}$ and are taken from \cite{AHearn1982}. CN and NH fluorescence efficiencies vary with both the heliocentric velocity and distance (because of the strong change in the number of rotational levels populated with heliocentric distance). CN and NH g-factors are taken from \cite{Schleicher2010} and from \cite{Meier1998}, respectively. The OH g-factor value of the 0-0 band centered near 3090 $\AA$ varies with the heliocentric velocity and whether or not the lambda-doublet ground state is quenched; scaled by r$_h^{-2}$ \citep{Schleicher1988}.

\begin{table}
	\begin{center}
		\caption{Scale lengths and life times of molecules at 1 au scaled by r$_h^{-2}$ \citep{A'Hearn1995}.}
		\begin{tabular}{lcccc} 
			\hline
			Molecules  &  Parent & Daughter & Parent  & Daughter     \\
			&  (km)     &  (km)  &  life time  &   life time  \\
			&           &        & (s)               & (s)\\
			\hline 
			OH(0,0) & 2.4 $\times$10$^4$ & 1.6 $\times$10$^5$ &  2.4 $\times$10$^4$ & 1.6 $\times$10$^5$ \\
			NH(0,0) & 5.0 $\times$10$^4$ & 1.5 $\times$10$^5$ &  5.0 $\times$10$^4$ & 1.5 $\times$10$^5$ \\
			CN($\Delta\upsilon$=0) & 1.3 $\times$10$^4$ & 2.1 $\times$10$^5$ & 1.3 $\times$10$^4$ & 2.1 $\times$10$^5$ \\
			C$_3$($\lambda$=4050 \AA{}) & 2.8 $\times$10$^3$ & 2.7 $\times$10$^5$ &  2.8 $\times$10$^3$ & 2.7 $\times$10$^5$ \\
			C$_2$($\Delta\upsilon$=0) & 2.2 $\times$10$^4$ & 6.6 $\times$10$^4$ & 2.2 $\times$10$^4$ & 6.6 $\times$10$^4$ \\
			\hline	
			\hline
			\label{tab1} 
		\end{tabular}
	\end{center}
\end{table}

\begin{figure}
	\includegraphics[scale=0.38]{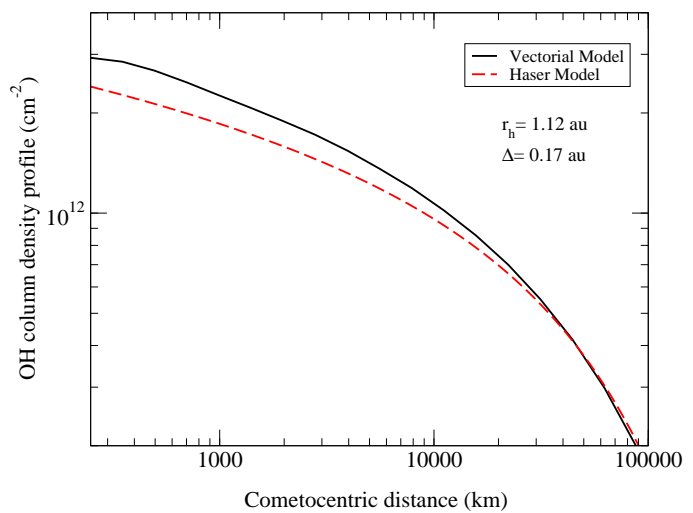}
	\caption{Comparison of the OH column density profile of 41P from the Vectorial and the Haser models for a production rate of 2.21$\times$10$^{27}$ molecules/s obtained on March 14, 2017.}
	\label{vect_model}
\end{figure}

In order to derive the production rates, we have converted the flux for different gas species (OH, NH, CN, C$_3$ and C$_2$) to column densities and we have adjusted their profiles with a Haser model \citep{Haser1957}. This rather simple model is based on a number of assumptions. Outgassing is assumed to be isotropic and the gas streams from the nucleus with a constant radial velocity. Parent molecules coming off the nucleus are decaying by photo dissociation to produce the observed daughter molecules. Table \ref{tab1} shows the scale lengths and life times of different daughter and parent molecules at 1 au from the Sun \citep{A'Hearn1995} that we used to adjust the radial profiles to the Haser model. The Vectorial model introduced by \cite{Festou1981a,Festou1981b} is physically more realistic, it considers the existence of collisions in the inner coma producing a non radial motion of the molecules. It becomes less pronounced when the cometocentric distance increases, and the release of energy when daughter molecules are produced may result in an additional non radial motion throughout the entire cometary coma. It is noted that a vectorial model appears to be required to represent the density distribution of some radicals in a comet. But the Vectorial model has more free parameters, which are usually poorly determined and may be a major source of uncertainty in the determination of gas production rates. Most of the authors are still using the Haser model to derive the production rates. In this work, we decided to use this model to compute the production rates in order to compare our results with others. Figure \ref{vect_model} shows the comparison of Vectorial and Haser Model for a similar OH production rate (Q(OH)=2.21$\times$10$^{27}$  molecules/s computed from a fit at 10000 km from nucleus when 41P was at 1.12 au from the Sun and at 0.17 au from Earth). The Vectorial model column density is retrieved from the following website\footnote{Vectorial model: \url{http://www.boulder.swri.edu/wvm-2011/}}. 
In order to derive the OH column density profile from the Vectorial model website, we set different parameters. So that, we used the same scale lengths and lifetimes for the Haser model given in Table \ref{tab1}. We set the velocities of the parent and daughter molecules at 1 km/s for a heliocentric distance of 1 au. The program internally adjusts the parent molecule velocity with the factor of $\sqrt(1/r_h)$, while the model considered that the daughter molecule velocity is independent of the heliocentric distance and no scaling for distance is made.

In order to estimate the dust production, we derived the Af$\rho$ parameter, quantity introduced by \cite{A'Hearn1984}, from the dust profiles using the narrow-band BC, GC and RC filters and the broad-band Rc filter. In the cometary coma, the brightness of the continuum is affected by the dependence on the solar phase angle of the scattering of sunlight by cometary grains. In all cases, there is a strong increase in the forward direction of scattering, and a much smaller peak at small backscattering angles. In practice these effects can reach up to a factor of 3 over a range of phase angles between 0$^\circ$ and 110$^\circ$, and much more at higher phase angles \citep{Schleicher1998}. An early phase function for cometary dust was constructed by \cite{Divine1981}, which appears to provide a reasonable match to comet observations between phase angles of $\sim$15$^\circ$-70$^\circ$. But this phase function is too shallow at smaller phase angles \citep{Ney1976,Hanner1989,Schleicher1998}, and does not increase sufficiently fast at large phase angles \citep{Marcus2007}. In this work, the A($\theta$)f$\rho$ values were corrected from the phase angle effect  according to the phase function normalized at $\theta$=0$^\circ$ given by \cite{Schleicher2010}\footnote{Phase function for comets: \url{http://asteroid.lowell.edu/comet/dustphase.html}}.

\section{Narrow-band photometry}
\label{sec3}

In this section, we present the narrow-band photometry with a description of the activity and composition evolution of comet 41P before and after perihelion and comet 45P after perihelion.

We adjust the Haser model \citep{Haser1957} on the observed radial profiles for the gas filter images. Since the comets presented in this work have been observed over a wide range of heliocentric and geocentric distances, we decided to make the adjustment at a physical distance around 10000 km from the nucleus, to derive the gas production rates. In most cases, the signal-to-noise ratio is still high at this cometocentric distance, and the influence of the seeing effect is limited to the first few pixels and the dust contamination is low.  In the dust filter images, the signal-to-noise is very weak at large cometocentric distance. We then decided to measure the Af$\rho$ values at 5000 km from the nucleus.
		
\subsection{Comet 41P/Tuttle-Giacobini-Kresak}

\begin{figure*}
	\centering
	\includegraphics[scale=0.6]{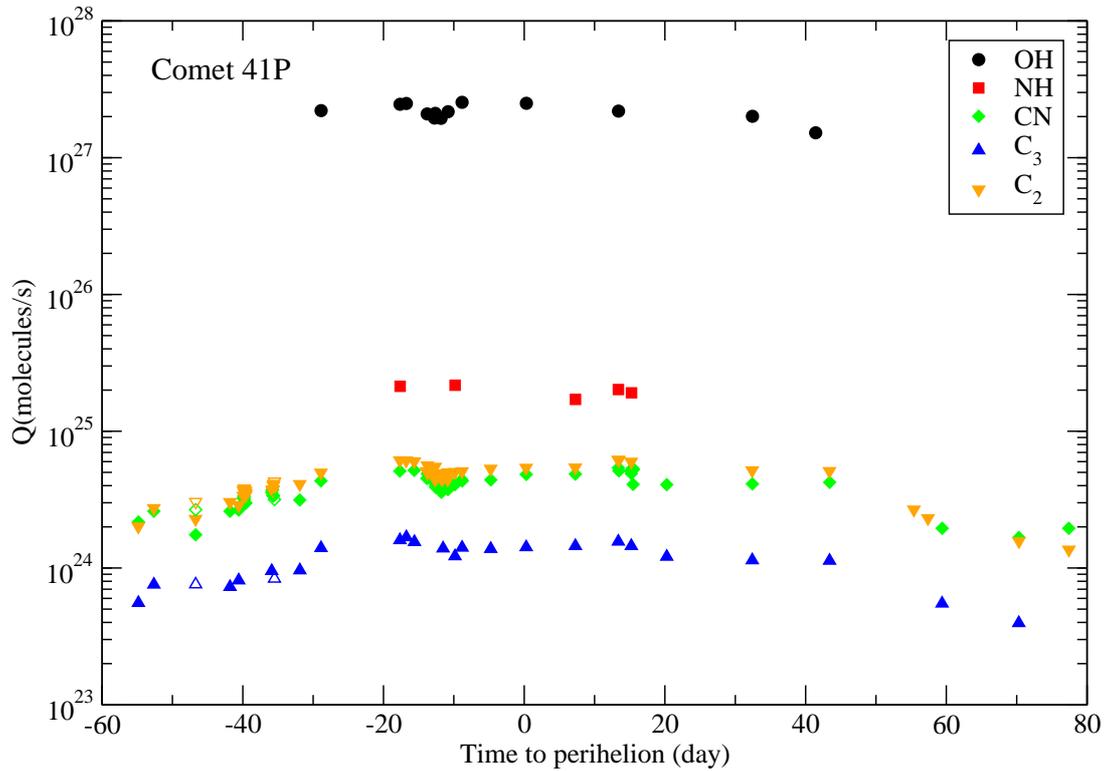}
	\caption{The production rates of comet 41P for each observed molecular species as a function of time to perihelion. The TN data are represented with filled symbols and TS data with open symbols. The values and their uncertainties are given in Table \ref{rate41P}.}
	\label{fig1}
\end{figure*}

The derived production rates for each gas species and A($\theta$)f$\rho$ values for the green, blue and red continuum are given in Table \ref{rate41P} with their errors. We started to detect CN, C$_2$ and C$_3$ species at the beginning of the second half of February, 2017 (r$_h$=1.27 au), when the OH and NH species were detected for the first time on March 11 and March 26, respectively. The gas production rates relatively increase towards the perihelion. We continued to monitor the comet 41P until the perihelion when the comet was at 1.05 au from the Sun. The OH, NH, CN, C$_2$ and C$_3$ production rates did not vary much between March 26 and May 27, 2017. We also monitored 41P after perihelion, the CN and C$_2$ were detected until July 20, when the comet was at 1.63 au from the Sun, and C$_3$ was not detected after June 22 (r$_h$=1.39 au). The CN, C$_2$ and C$_3$ production rates start to decrease from May 27 until the end of monitoring. OH was detected until May 24 (r$_h$=1.18 au) with a production rate of (1.57$\pm$0.08)$\times$10$^{27}$ and NH reached (1.91$\pm$0.10)$\times$10$^{25}$  molecules/s on April 27. Figure \ref{fig1} shows the production rates for all gas species in logarithmic scale as a function of time to perihelion. The OH, CN, C$_3$ and C$_2$ production rates from TN are in agreement with the few measurements obtained with TS at the beginning of the monitoring.

\begin{table*}
	\begin{center}
		\caption{OH, NH, CN, C$_2$ and C$_3$ production rates and A($\theta$=0$^\circ$)f$\rho$ parameter for comet 41P/Tuttle–Giacobini–Kresák}
		\resizebox{\textwidth}{!}{%
			\begin{tabular}{lccccccccccc}
				\hline  
				\hline
				UT Date    & $r_h$ & $\bigtriangleup$  &  \multicolumn{5}{c}{Production rates  (10$^{24}$molecules/s)}   &  \multicolumn{4}{c}{A($\theta$=0$^\circ$)f$\rho$ (cm)} \\
				&(au) & (au)  & Q(OH)  & Q(CN)	& Q(NH)	& Q(C$_2$) & Q(C$_3$) & BC & RC & GC & Rc \\
				\hline 
				\hline
				2017 Feb 16.88&1.27&0.29&  & 2.20$\pm$0.04 &  & 2.26$\pm$0.03 & 0.61$\pm$0.02 &31.0$\pm$1.4 & 39.5$\pm$0.7& 29.7$\pm$1.6&36.0$\pm$0.2  \\
				2017 Feb 19.08&1.26&0.27&  & 2.55$\pm$0.04 &  & 2.66$\pm$0.03 & 0.73$\pm$0.02 &31.9$\pm$1.8 & 43.7$\pm$1.3& 37.0$\pm$0.7&42.8$\pm$0.2  \\
				2017 Feb 25.04&1.22&0.24& &  & 				 & 2.51$\pm$0.03 & 0.67$\pm$0.01 &29.8$\pm$1.2 & 39.9$\pm$1.2& &41.3$\pm$0.1 \\
				2017 Feb 25.10$^\dagger$&1.22&0.24& &2.67$\pm$0.02  & 	& 3.04$\pm$0.03& 0.76$\pm$0.03 &25.1$\pm$1.1 & 37.9$\pm$0.5& &46.2$\pm$0.3 \\
				2017 Mar 01.92&1.19&0.22& & 2.63$\pm$0.05 &  & 3.14$\pm$0.05 & 0.74$\pm$0.02 &45.3$\pm$2.3 & &  &55.3$\pm$0.2\\
				2017 Mar 03.12&1.18&0.21&  & 2.69$\pm$0.03 &                & 3.16$\pm$0.02 & 0.87$\pm$0.02 &36.4$\pm$2.1 & 49.3$\pm$1.2& &55.8$\pm$0.6 \\
				2017 Mar 03.84&1.18&0.21& & 3.35$\pm$0.05 & 				 & 3.94$\pm$0.05& & & &&  \\
				2017 Mar 04.16&1.17&0.20& 			   & 3.02$\pm$0.05 & 				 & 3.78$\pm$0.05& & & & & \\
				2017 Mar 07.84&1.15&0.19&  & 3.55$\pm$0.08 &        		 & 4.36$\pm$0.05&0.88$\pm$0.05&50.6$\pm$5.1&66.8$\pm$3.6& &74.0$\pm$2.6 \\
				2017 Mar 08.12&1.15&0.19& 			   & 3.35$\pm$0.05 & 				 & 3.24$\pm$0.04& & & & &79.1$\pm$2.0 \\
				2017 Mar 08.12$^\dagger$&1.15&0.19& 	& 3.16$\pm$0.03 &  & 4.29$\pm$0.06& 0.84$\pm$0.07 & &50.9$\pm$1.9 & &70.6$\pm$1.7 \\
				2017 Mar 11.84&1.13&0.18& 2190$\pm$127 & 3.11$\pm$0.06 & 				 & 3.83$\pm$0.05& 0.87$\pm$0.04 &56.1$\pm$3.9 &64.5$\pm$8.3 &&69.0$\pm$5.3 \\
				2017 Mar 14.84&1.12&0.17& 2270$\pm$155 & 4.34$\pm$0.06 & 				 & 5.10$\pm$0.06& 1.38$\pm$0.02 &54.7$\pm$1.4 & & &87.3$\pm$1.1 \\
				2017 Mar 26.04&1.07&0.15& 2530$\pm$140 & 5.07$\pm$0.09 & 21.10$\pm$0.64 & 6.07$\pm$0.09& 1.56$\pm$0.02 &54.8$\pm$1.2 &73.7$\pm$2.9 & 63.4$\pm$1.5&93.0$\pm$1.8 \\
				2017 Mar 26.96&1.07&0.14& 2570$\pm$128 &  & 				 & 6.21$\pm$0.10& 1.67$\pm$0.02 &56.2$\pm$2.4 &71.7$\pm$2.7 & &89.5$\pm$1.2 \\
				2017 Mar 28.08&1.07&0.14&  & 5.17$\pm$0.07 & 				 & 6.03$\pm$0.08& 1.53$\pm$0.01 &53.5$\pm$1.8 &71.4$\pm$2.6 & &89.5$\pm$1.4 \\
				2017 Mar 29.92&1.06&0.14& 				& 4.54$\pm$0.09 & 				 & 5.21$\pm$0.07& &48.1$\pm$1.7 & & & \\
				2017 Mar 29.96&1.06&0.14& 2150$\pm$146 & 4.93$\pm$0.08 & 				 & 5.50$\pm$0.07& & & & &81.1$\pm$1.1 \\
				2017 Mar 30.00&1.06&0.14& 				& 4.88$\pm$0.09 & 				 & 4.92$\pm$0.08& &49.5$\pm$2.0 & & & \\
				2017 Mar 30.96&1.06&0.14& 2010$\pm$131 & 4.21$\pm$0.10 & 				 & 5.00$\pm$0.09& &46.9$\pm$1.7 & & &71.1$\pm$1.4 \\
				2017 Mar 31.00&1.06&0.14& 				& 4.41$\pm$0.09 & 				 & 4.52$\pm$0.09& & & & & \\
				2017 Mar 31.04&1.06&0.14& 2180$\pm$144 & 4.08$\pm$0.07 & 				 & 4.29$\pm$0.08& & & & & \\
				2017 Mar 31.12&1.06&0.14& 				& 3.94$\pm$0.07 & 				 & 4.66$\pm$0.08& & & & & \\
				2017 Mar 31.16&1.06&0.14& 2470$\pm$199 & 3.91$\pm$0.07 & 				 & 4.28$\pm$0.07& &41.4$\pm$2.3 & &&71.3$\pm$1.2  \\
				2017 Mar 31.88&1.06&0.14& 2010$\pm$173 & 3.80$\pm$0.06 & 				 & 4.69$\pm$0.06& &34.0$\pm$3.1 & & &67.3$\pm$0.9 \\
				2017 Mar 31.96&1.06&0.14& 				& 3.67$\pm$0.09 & 				 & 4.44$\pm$0.09& & &50.6$\pm$1.4 & &63.7$\pm$1.1 \\
				2017 Apr 01.00&1.06&0.14& 				& 3.83$\pm$0.09 & 				 & 4.52$\pm$0.09& & & & 43.7$\pm$2.0 &65.7$\pm$1.5 \\
				2017 Apr 01.04&1.06&0.14& 				& 3.58$\pm$0.07 & 15.10$\pm$0.63 & 4.55$\pm$0.10& & & & &64.8$\pm$1.7 \\
				2017 Apr 01.08&1.06&0.14& 				& 3.75$\pm$0.08 & 				 & 4.49$\pm$0.08& & &54.2$\pm$1.1 & &68.5$\pm$1.0 \\
				2017 Apr 01.16&1.06&0.14& 				& 3.91$\pm$0.07 & 				 & 4.54$\pm$0.08& 1.33$\pm$0.01 &41.9$\pm$1.1 & & & \\
				2017 Apr 01.88&1.06&0.14& 2250$\pm$173 & 3.77$\pm$0.06 & 				 & 4.43$\pm$0.06& &33.2$\pm$2.3 & & &59.7$\pm$1.2 \\
				2017 Apr 01.92&1.06&0.14&				& 4.05$\pm$0.06 & 				 & 4.77$\pm$0.06& &33.5$\pm$1.4 & & &64.5$\pm$1.6 \\
				2017 Apr 02.88&1.05&0.14& 				& 4.07$\pm$0.07 & 21.70$\pm$0.80 & 5.02$\pm$0.07& 1.23$\pm$0.02 &34.7$\pm$2.5 &49.6$\pm$0.8 & 41.1$\pm$1.1&60.5$\pm$1.0 \\
				2017 Apr 03.88&1.05&0.14& 2610$\pm$149 & 4.24$\pm$0.05 & 				 & 5.15$\pm$0.06& 1.38$\pm$0.02 &40.8$\pm$2.0 &53.7$\pm$2.8 &  &68.9$\pm$0.6\\
				2017 Apr 07.96&1.05&0.15&              &               & 				 & 5.42$\pm$0.08& 1.36$\pm$0.02 &31.6$\pm$3.8 &52.9$\pm$2.5 & &66.3$\pm$2.8\\
				2017 Apr 08.00&1.05&0.15& 2350$\pm$178 & 4.42$\pm$0.09 & 				 & & & & & & \\
				2017 Apr 13.00&1.05&0.15& 2590$\pm$157 & 4.84$\pm$0.10 & 				 & 5.41$\pm$0.07& 1.42$\pm$0.03 &34.3$\pm$4.5 &56.1$\pm$1.9 &  &66.1$\pm$1.9\\
				2017 Apr 20.00&1.05&0.17& 1840$\pm$131 & 4.81$\pm$0.07 & 17.00$\pm$0.78 & 5.24$\pm$0.07& 1.37$\pm$0.01 &36.7$\pm$1.1 & & 40.4$\pm$1.3& 60.6$\pm$1.2\\
				2017 Apr 26.08&1.06&0.18&              &               &                & 6.05$\pm$0.08& 1.50$\pm$0.02 &45.8$\pm$2.2 & &  &74.2$\pm$1.3 \\
				2017 Apr 26.12&1.06&0.18& 2260$\pm$54& 5.34$\pm$0.08 & 20.20$\pm$0.84 & & & & & & \\
				2017 Apr 26.21&1.06&0.18& 				& 5.12$\pm$0.06 & 				 & & & & & &72.5$\pm$1.4 \\
				2017 Apr 27.96&1.07&0.18& 				& 5.02$\pm$0.08 & 19.10$\pm$1.02 &5.81$\pm$0.07 & 1.38$\pm$0.02 &45.4$\pm$1.9 & & &72.9$\pm$1.4 \\
				2017 Apr 28.00&1.07&0.18& 2080$\pm$134 & 5.13$\pm$0.07 & 				 & & & & & & \\
				2017 Apr 28.21&1.07&0.19&			    & 5.15$\pm$0.06 & 				 & & & & & & \\
				2017 May 03.00&1.08&0.20& & 4.04$\pm$0.05 & 				 & 4.55$\pm$0.06& 1.18$\pm$0.02 &37.7$\pm$1.7 & & &56.2$\pm$1.0 \\
				2017 May 15.16&1.13&0.23& 2070$\pm$115 & 4.08$\pm$0.06 & 				 & 5.02$\pm$0.06& 1.08$\pm$0.03 &41.7$\pm$7.1 & & & \\
				2017 May 24.12&1.18&0.26& 1570$\pm$83& 3.39$\pm$0.03 &  & 3.94$\pm$0.03& 1.00$\pm$0.03 &42.3$\pm$2.3 & & & \\
				2017 May 26.12&1.20&0.27& & 4.23$\pm$0.05 &  & 5.19$\pm$0.04& 1.13$\pm$0.02 &55.0$\pm$3.1 & & & \\
				2017 Jun 07.12&1.28&0.32& 				&				&				 & 2.68$\pm$0.07&				  &33.2$\pm$2.9 & & &54.6$\pm$1.2 \\
				2017 Jun 09.12&1.29&0.33& 				&				&				 & 2.10$\pm$0.11&				  & & &	&48.3$\pm$5.1\\
				2017 Jun 10.04&1.30&0.33&				& 1.15$\pm$0.10 & 				 & & 0.94$\pm$0.03 & & & &44.6$\pm$6.7 \\
				2017 Jun 11.12&1.31&0.34& & 1.81$\pm$0.04 &  & & 0.23$\pm$0.08 & & & &38.5$\pm$5.2 \\
				2017 Jun 22.04&1.39&0.40&  & 1.69$\pm$0.03 &  & 1.81$\pm$0.04& 0.44$\pm$0.02 &21.9$\pm$1.5 & & &34.1$\pm$0.8 \\
				2017 Jun 29.12&1.45&0.46& 				& 2.03$\pm$0.04 & 				 & 2.00$\pm$0.04& & & & &36.2$\pm$0.5 \\
				2017 July 20.88&1.63&0.67 &  & 1.32$\pm$0.04 &  & 1.33$\pm$0.05& & & &  &20.5$\pm$2.3\\
				
				\hline	
				\hline 
				\label{rate41P}
			\end{tabular}}
		\end{center}
		\tablefoot{$r_h$ and $\bigtriangleup$ are the heliocentric and geocentric distances, respectively. ($^\dagger$) present the TRAPPIST-South measurements. The A(0)f$\rho$ values are printed at 5000 km from the comet's nucleus.}
	\end{table*}

\begin{figure*}
	\centering
	\includegraphics[scale=0.6]{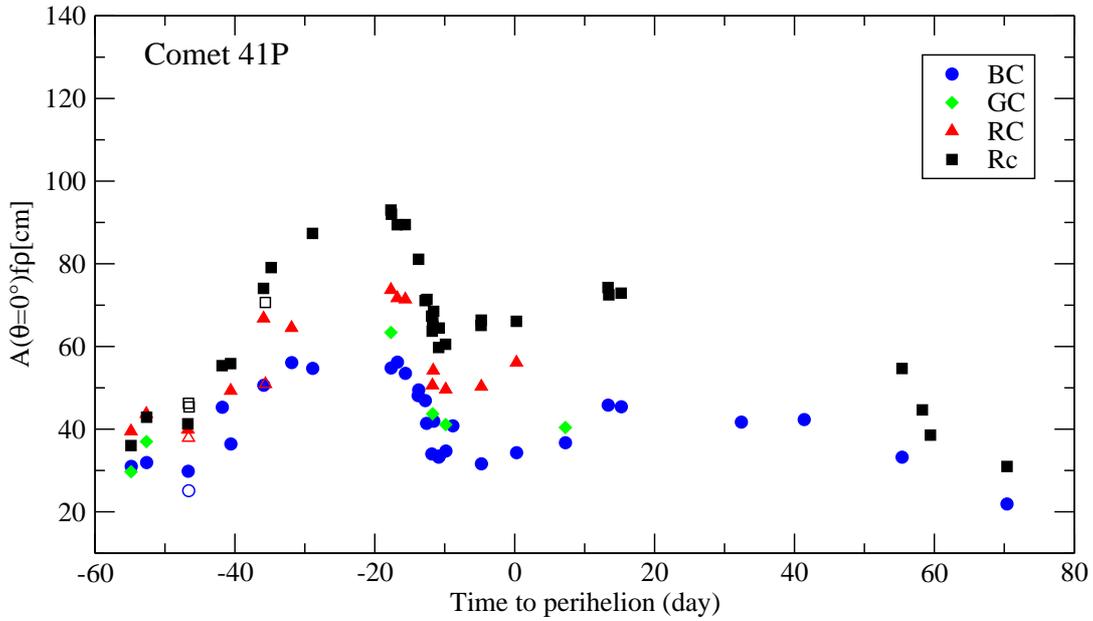}
	\caption{The A($\theta$=0$^\circ$)f$\rho$ parameter of comet 41P for the narrow-band filters (BC, RC and GC) and for the broad-band filter Rc as a function of time to perihelion. We normalized the A($\theta$)f$\rho$ values at 0$^\circ$ phase angle. The TN data are represented with filled symbols and TS data with open symbols. The values and their uncertainties are given in the Table \ref{rate41P}.}
	\label{fig2}
\end{figure*}

In order to estimate the dust activity, we observed 41P with the BC, RC and GC continuum filters in addition to the ones regularly performed with the broad-band Rc filter. The values of the A($\theta$)f$\rho$ parameter corrected from the phase angle are shown in Figure \ref{fig2} in a logarithmic scale as a function of time from perihelion. The present paper focuses on the analysis of the activity and the gas composition. The analysis of the dust coma of 41P has been done in \cite{Pozuelos2018}. They performed Monte Carlo simulations to characterize the dust environment as a function of the heliocentric distance using the model described in \cite{Moreno2012}. The total amount of dust produced was roughly 7.5$\times$10$^8$ kg, with a peak of activity of 110 kg/s a few days before perihelion. They concluded that 41P is a dust poor comet compared to other JFCs.

\begin{figure*}
	\centering
	\includegraphics[scale=0.6]{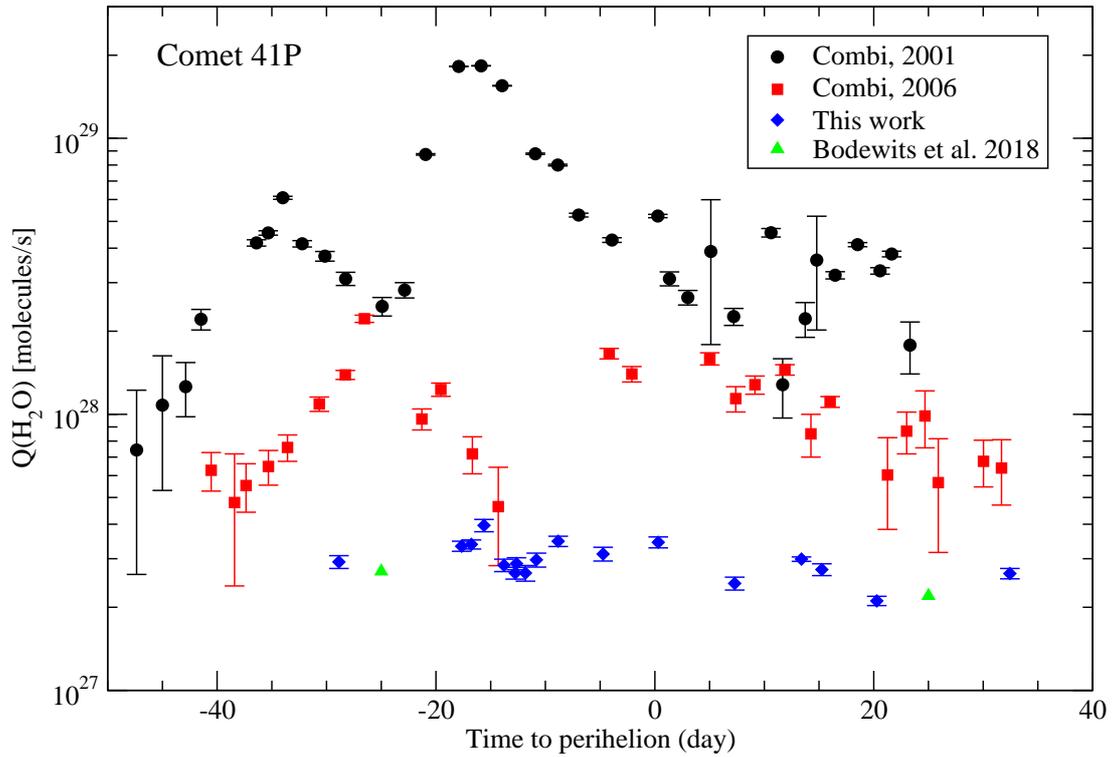}
	\caption{The logarithm of the water production rate for different apparitions of comet 41P in 2001, 2006 and 2017 as a function of time to perihelion. \cite{Combi2017} derived the production rates from hydrogen Lyman-$\alpha$ emission observed by the SWAN instrument on board SOHO in 2001 (black circles) and 2006 (red squares). Our water production rates (blue diamonds) were derived from the OH production rates using the relationship given in \cite{Cochran1993}. The water production rates given in \cite{Bodewits2018} are derived from the Swift/UVOT observations of OH emissions (green triangles) and are in good agreement with the TRAPPIST ones.}
	\label{fig3}
\end{figure*}

Figure \ref{fig3} presents the comparison of the water production rates of 41P at different apparitions as a function of time to perihelion. The water production rates, corresponding to the apparition in 2001 (black circles) and 2006 (red squares), were derived from the hydrogen Lyman-$\alpha$ emission observed with the SWAN instrument on board the Solar and Heliospheric Observatory (SOHO) spacecraft. For the SWAN data, 1$\sigma$ stochastic errors are shown; systematic uncertainties are at the 30\% level \citep{Combi2006,Combi2011}. \cite{Bodewits2018} used recently Swift/UVOT observations of hydroxyl (OH) emission to determine the water production rate in 2017 (green triangles). These values are in good agreement with our TRAPPIST water production rates. Using our data, we computed a  vectorial-equivalent water production rates (blue diamonds) from our Haser-model OH production rates using Q(H$_2$O)=1.361$\times$r$_h^{-0.5}\times$Q(OH) given in \cite{Cochran1993}. The water production rate peaked around (183$\pm$0.26)$\times$10$^{27}$ molecules/s on December 21, 2000 when the comet was at 1.07 au from the Sun and (22.2$\pm$0.67)$\times$10$^{27}$ molecules/s on May 15, 2006 when 41P was at 1.10 au from the Sun \citep{Combi2017}. As shown in Figure \ref{fig3}, two outbursts have been detected at optical wavelengths as peaks 33 and 15 days before perihelion in 2001 \citep{Kronk2017,Combi2017}. However, no outburst was detected during the 2017 perihelion passage. From the data shown in Figure \ref{fig3}, we conclude that there was a $\sim$3.5$\times$ decrease in water outgassing between 2001 and 2006, for two continuous orbits ($\sim$5 years), and a $\sim$5$\times$ decrease between 2006 and 2017, an interval of two orbits. This implies a relatively consistent drop of about 30\% to 40\% from one apparition to the next. Generally, the distribution of active surface areas of comets correlate with the smaller nuclei \citep{A'Hearn1995}. As 41P has a small nucleus, \cite{Bodewits2018} claims that more than 50\% of the surface of 41P could be active, while most comets have usually less than 3\% of their surface active \citep{A'Hearn1995}. We note that the water production rates are coming from different techniques and observations at different wavelengths, which means the comparison of the results should be taken with caution.

\subsection{Comet 45P/Honda–Mrkos–Pajdušáková}  

\begin{figure*}
	\centering
	\includegraphics[scale=0.6]{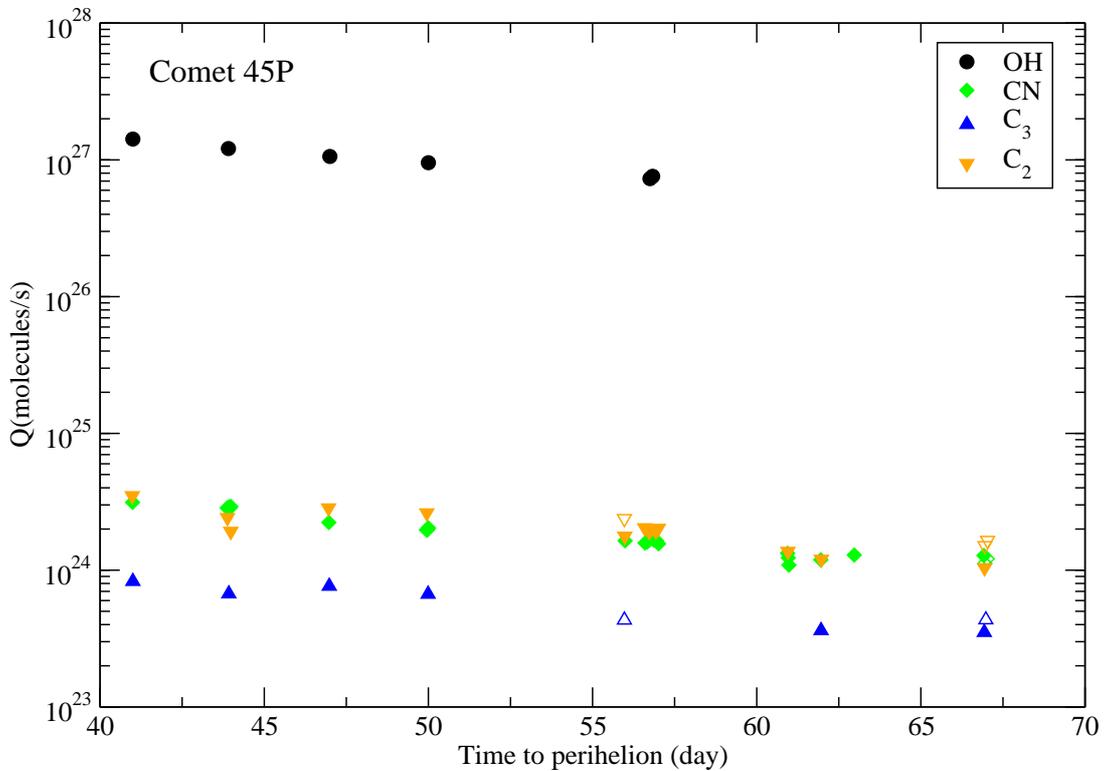}
	\caption{The logarithm of the production rates of comet 45P for each observed molecular species as a function of time to perihelion. The TN data are represented with filled symbols and TS data with open symbols. The values and their uncertainties are given in Table \ref{rate45P}. }
	\label{fig45P_gases}
\end{figure*}

\begin{figure*}
		\centering
		\includegraphics[scale=0.63]{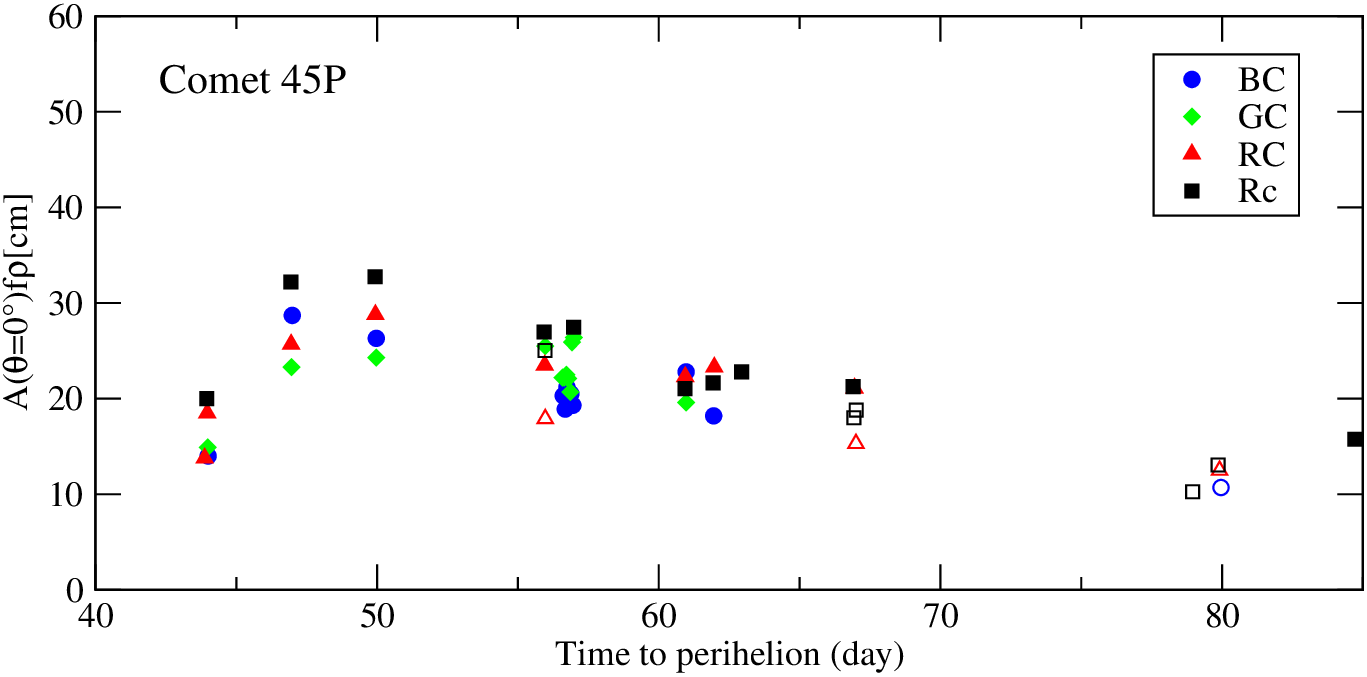}
		\caption{The A($\theta$=0$^\circ$)f$\rho$ parameter of comet 45P for the narrow-band filters (BC, RC and GC) and for the broad-band filter Rc as a function of time to perihelion. We normalized the A($\theta$)f$\rho$ values at 0$^\circ$ phase angle. The TN data are represented with filled symbols and TS data with open symbols. The values and their uncertainties are given in Table \ref{rate45P}.}
		\label{fig45P_Afp}
	\end{figure*}

The mean OH, CN, C$_2$ and C$_3$ production rates that we obtain for 45P during our observations, were (1.07$\pm$0.46)$\times$10$^{27}$, (1.77$\pm$0.05)$\times$10$^{24}$, (1.93$\pm$0.06)$\times$10$^{24}$, and (0.65$\pm$0.04)$\times$10$^{24}$  molecules/s, respectively. While NH was not detected, the OH, CN, C$_2$ and C$_3$ production rates and Af$\rho$ in the broad- and narrow-band filters  are given with their uncertainties in Table \ref{rate45P}. The evolution of comet 45P gas and dust activity as a function of time to perihelion is summarized in Figure \ref{fig45P_gases} and in Figure \ref{fig45P_Afp}, respectively. As shown in Figure \ref{fig45P_gases}, the gas species production rates decrease after perihelion (from 0.97 au to 1.62 au from the Sun). The lack of data for this comet does not allow us to compute the production rates slopes, but it still shows a dependence on heliocentric distance. The A($\theta$)f$\rho$ parameter, corrected from the phase angle effect, shows a peak of 33 cm in Rc filter around 50 days after perihelion. Then, the  A(0)f$\rho$ starts to decrease with heliocentric distance. 45P phase angle changes from 110$^\circ$ to 20$^\circ$ during the entire observing period. The TRAPPIST-North and -South results are again in agreement.

Comet 45P was observed by numerous observers at different wavelengths so as to detect the parent molecules (H$_2$O, NH$_3$, HCN and C$_2$H$_2$). \cite{DiSanti2017} reported various molecules production rates of 45P on 6-8 January, 2017 (r$_h$=0.55 au, one week after perihelion), using the high spectral resolution cross-dispersed facility, iSHELL, at the NASA Infrared Telescope Facility on Maunakea. The authors obtained a mean value of the water production rate of 1.5$\times$10$^{28}$ molecules/s noting a decrease during these two days. We have a mean water production rate of 1.5$\times$10$^{27}$ molecules/s. Possible causes for this discrepancy could be the different size of the field of view and also the technique used to derive the water production rate. Moreover, \citep{DiSanti2017} had observed 45P around the perihelion, while we started to observe this comet 40 days after perihelion. Using radar observations made on February 9, 2017, \cite{Lejoly2017} detected a skirt of material around the nucleus of the comet, indicating that there is a large population of centimeter-sized grains being emitted from the comet.
	
45P was also observed in the past. During its passage in July 1990, it was observed with the Lowell Observatory when the comet was at 1.15 au from the Sun and 0.36 au from the Earth. The production rates obtained for 5 gas species were:  Q(OH)=6.16$\times$10$^{26}$, Q(NH)=3.80$\times$10$^{24}$, Q(CN)=8.12$\times$10$^{23}$, Q(C$_2$)=9.12$\times$10$^{22}$, Q(C$_3$)=1.38$\times$10$^{24}$ molecules/s and the Af$\rho$ parameter was 11 cm measured at 5000 km from the nucleus. At this time, the water production rate was about 8$\times$10$^{26}$ molecules/s \citep{Randall1992,Osip1992,A'Hearn1995}. The comparison of these results with our mean production rate values of 45P in 2017 shows that OH and CN production rates increased by a factor 2 and C$_2 $ increased by 20, while the production rate of C$_3$ decreased by 2. We conclude that 45P has developed its activity between 1999 and 2017 passages.

	\begin{table*}[!]
		\begin{center}
			\caption{OH, CN, C$_2$ and C$_3$ production rates and A($\theta$=0$^\circ$)f$\rho$ parameter for comet 45P/Honda-Mrkos-Pajdusakova }
			\resizebox{\textwidth}{!}{%
				\begin{tabular}{lcccccccccc}
					\hline  
					\hline
					UT Date    & $r_h$ & $\bigtriangleup$  &  \multicolumn{4}{c}{Production rates  (10$^{24}$ molecules/s)}   &  \multicolumn{4}{c}{A($\theta$=0$^\circ$)f$\rho$ (cm)} \\
					&(au) & (au)  & Q(OH)  & Q(CN)	& Q(C$_2$) & Q(C$_3$) & BC & RC & GC &Rc \\
					\hline 
					\hline
					2017 Feb 10.25  &0.97  &0.08  &1420$\pm$51 &3.13$\pm$0.13 & 3.50$\pm$0.12  &0.82$\pm$ 0.04  &   &   &  & \\	
					2017 Feb 13.12  &1.01  &0.09  & &2.85$\pm$0.07 &2.42$\pm$0.10   &  &   &13.8$\pm$2.2   &  &  \\
					2017 Feb 13.16  &1.01  &0.09  &1210$\pm$55 &2.85$\pm$0.06 &   &0.67$\pm$0.04  &   &   &  &  \\
					2017 Feb 13.20  &1.01  &0.09  & &2.89$\pm$0.04 &1.92$\pm$0.08  &  &   &18.5$\pm$1.6   &14.9$\pm$2.0  &20.0$\pm$2.0  \\
					2017 Feb 16.20  &1.05  &0.11  &1060$\pm$43 &2.23$\pm$0.06 &2.85$\pm$0.10 &0.76$\pm$0.02  &14.0$\pm$2.5   &25.7$\pm$1.9   &23.3$\pm$2.2  &32.2$\pm$1.8  \\
					2017 Feb 19.16  &1.09  &0.13  & &1.97$\pm$0.05 &2.62$\pm$0.08   &  &   &28.8$\pm$1.6   &  & 32.7$\pm$1.4 \\
					2017 Feb 19.20  &1.09  &0.13  & &1.98$\pm$0.05 &   &0.67$\pm$0.02  &28.7$\pm$2.0   &   &24.3$\pm$1.5  &  \\
					2017 Feb 19.25  &1.09  &0.13  &953$\pm$52 &2.04$\pm$0.07 &   &  &   &   &  &  \\
					2017 Feb 25.20  &1.18  &0.20  & &1.64$\pm$0.04 & 1.77$\pm$0.05   &  &   &23.5$\pm$1.0   &25.5$\pm$1.6  & 26.9$\pm$0.8  \\
					2017 Feb 25.25$^\dagger$  &1.18  &0.20  & & &2.38$\pm$0.02 &0.43$\pm$0.04 & &17.9$\pm$1.7  &  &25.0$\pm$0.9 \\
					2017 Feb 25.84  &1.19  &0.21  & &1.58$\pm$0.03 &2.04$\pm$0.04   &  &26.3$\pm$2.3   &   &22.2$\pm$1.5  & \\
					2017 Feb 25.92  &1.19  &0.21  & &1.61$\pm$0.03 & 1.90$\pm$0.05  &  &   &   &22.2$\pm$1.1  &  \\
					2017 Feb 25.96  &1.19  &0.21  &744$\pm$32 &1.66$\pm$0.03 & 1.91$\pm$0.04  &  & 20.3$\pm$1.3  &   &22.5$\pm$1.3  &  \\
					2017 Feb 26.12  &1.19  &0.22  & &1.66$\pm$0.03 &2.10$\pm$0.05   &  & 21.2$\pm$1.5  &   &22.1$\pm$1.3  &   \\
					2017 Feb 26.16  &1.19  &0.22  & &1.65$\pm$0.03 &1.93$\pm$0.05   &  &20.2$\pm$1.1   &   &25.9$\pm$1.5  &  \\
					2017 Feb 26.25  &1.19  &0.22  & &1.56$\pm$0.16 &2.02$\pm$0.07   &  &   &   & 26.4$\pm$1.9 &27.5$\pm$0.8 \\
					2017 Mar 02.16  &1.25  &0.27  & &1.33$\pm$0.03 &1.37$\pm$0.04   &  &   &   &  & 21.0$\pm$0.6 \\
					2017 Mar 02.20  &1.25  &0.27  & &1.16$\pm$0.04 &   &  &19.3$\pm$1.5   &22.3$\pm$1.0   & 19.6$\pm$1.5 &  \\
					2017 Mar 03.16  &1.26  &0.28  & &1.19$\pm$0.04 &1.20$\pm$0.04   &0.36$\pm$0.03  &22.8$\pm$2.7   & 23.3$\pm$0.7  &  &21.6$\pm$0.6  \\
					2017 Mar 04.20  &1.28  &0.30  & &1.29$\pm$0.04 &   &  &   &   &  &22.8$\pm$0.6  \\
					2017 Mar 08.16  &1.33  &0.35  & &1.28$\pm$0.03 &1.04$\pm$0.04  & 0.35$\pm$0.03 & 18.2$\pm$2.0  &21.1$\pm$2.0   & &21.2$\pm$0.7  \\
					2017 Mar 08.16$^\dagger$  &1.33  &0.35  & &1.12$\pm$0.05 &  &  &   &15.3$\pm$1.3   &  &18.0$\pm$1.2  \\
					2017 Mar 08.25$^\dagger$  &1.33  &0.36  & &1.21$\pm$0.06 &1.58$\pm$0.07  &0.44$\pm$0.13  &   &   &  &18.8$\pm$1.2\\		
					2017 Mar 20.16$^\dagger$  &1.49  &0.54  & &1.01$\pm$0.03 &1.09$\pm$0.06  &  &  &   &  &10.3$\pm$1.0  \\		
					2017 Mar 21.16$^\dagger$  &1.50  &0.56  & & &1.10$\pm$0.10  &  &10.7$\pm$1.9   & 12.5$\pm$1.7  &  &13.0$\pm$0.9  \\
					2017 Mar 25.94 & 1.56  & 0.36 & & & & & & & &15.7$\pm$0.3 \\	
					\hline	
					\hline 
					\label{rate45P}
				\end{tabular}}
			\end{center}
			\tablefoot{$r_h$ and $\bigtriangleup$ are the heliocentric and geocentric distances, respectively. ($^\dagger$) present the TRAPPIST-South measurements. The A(0)f$\rho$ values are printed at 5000 km from the comet's nucleus.}
		\end{table*}

\subsection{Production rates ratios and comparison to other comets}
\setlength{\tabcolsep}{4pt}
\begin{table}
	\begin{center}
		\caption{Mean of the logarithm of production rates and A(0)f$\rho$ ratios with respect to OH, and for C$_2$ with respect to CN for 41P and 45P compared to the mean values presented in \cite{A'Hearn1995} and \cite{Schleicher2008}.}
			{\small 
		\begin{tabular}{lcccc}
			\hline  
			\hline
			& \multicolumn{4}{c}{Log Production rate ratio } \\
			\cline{2-5}
			Species &41P\tablefootmark{a} & 45P\tablefootmark{b}	&  \textcolor{blue}{Schleicher}  & \textcolor{blue}{A'Hearn}    \\
			  &        &    &  (\textcolor{blue}{2008})\tablefootmark{c}  & \textcolor{blue}{et al.} (\textcolor{blue}{1995})\tablefootmark{d} \\
			\hline 
			C$_2$/CN &0.06$\pm$0.003 & 0.04$\pm$0.003 &  0.10 &0.06$\pm$0.10  \\
			C$_2$/OH &-2.64$\pm$0.04 &-2.75$\pm$0.01  & -2.46 &-2.44$\pm$0.20   \\
			CN/OH    &-2.70$\pm$0.04 &-2.78$\pm$0.01  & -2.55 &-2.50$\pm$0.18   \\
			C$_3$/OH &-3.25$\pm$0.03 &-3.21$\pm$0.02  & -3.12 &-3.55$\pm$0.29   \\
			NH/OH    &-2.03$\pm$0.01 & --  & -2.23 &-2.37$\pm$0.27    \\
			Af$\rho$/OH &-25.70$\pm$0.03  &-25.71$\pm$0.34 & -- &-25.80$\pm$0.40 \\	
			\hline	
			\hline
			\label{ratios_45P}  
		\end{tabular}}
		\vspace{-0.4cm}
		\tablefoot{The A($\theta$=0$^\circ$)f$\rho$ values used for 41P and 45P are derived from BC filter. The ratio of A(0)f$\rho$ to Q(OH) has units of cm s /molecules.\\
			\tablefoottext{a}{the mean values computed for pre- and post-perihelion.}\\
			\tablefoottext{b}{the mean values computed for post-perihelion.}\\
			\tablefoottext{c}{the mean values for typical comets given in \cite{Schleicher2008}.}\\
			\tablefoottext{d}{the mean values for typical comets given in \cite{A'Hearn1995}.}}
	\end{center}
\end{table}
	
\cite{A'Hearn1995} presented a chemical classification for 85 comets observed between 1976 and 1992 using production rate ratios with respect to CN and to OH. In order to compare 41P and 45P to this taxonomy, we computed those ratios for all the gas species. Our results show that 41P and 45P are "typical" comets of the JF. The typical comets are defined in \cite{A'Hearn1995} as those having log[Q(C$_2$)/Q(CN)]$\geq$ -0.18, which was updated by \cite{Schleicher2006} to -0.11. The mean value of this quantity during our monitoring of 41P and 45P are 0.06$\pm$0.003 and 0.04$\pm$0.003, respectively, in agreement with the mean value given in \cite{A'Hearn1995} and \cite{Schleicher2008} for JFCs (see Table \ref{ratios_45P}). Preliminary results from the Near Infrared Spectrograph indicate that 41P has typical C$_2$H$_2$ and HCN abundances compared to other JFCs, while the C$_2$H$_6$ abundance is similar to that of Nearly Isotropic Comets, but is enriched compared to other JFCs \citep{McKay2018}.

	\begin{figure}
		\includegraphics[scale=0.41]{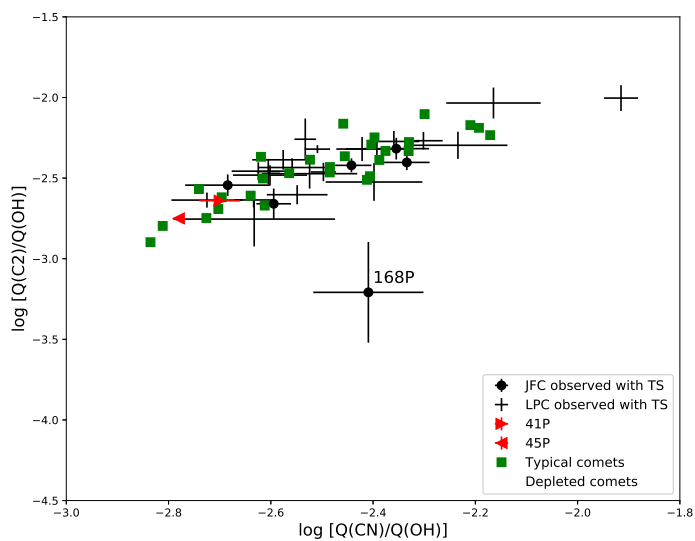}
		\caption{The logarithm of the ratio of C$_2$ to OH production rates as a function of the logarithm of the ratio of CN to OH, for comets 41P (triangle right) and 45P (triangle left) compared to JFCs (filled circles) and Long Period Comets (open circles) observed with TS between 2010 and 2016 \citep{Opitom2016thesis}, and also with typical (filled squares) and depleted comets (open squares) given in \cite{A'Hearn1995}.}
		\label{Ahearn}
	\end{figure}
	
\begin{figure*}
	\resizebox{\hsize}{!}{
		\includegraphics[scale=0.76]{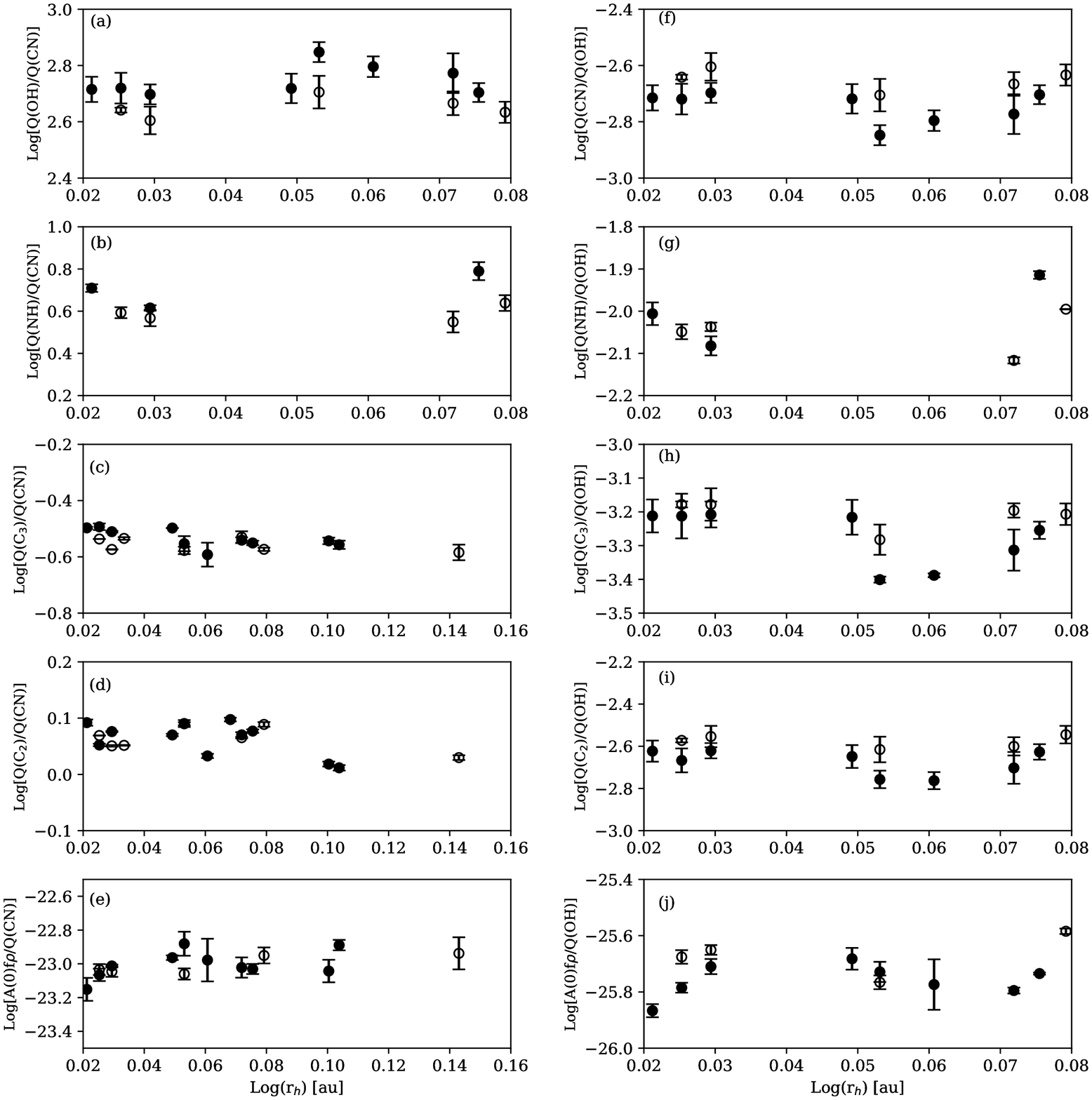}}
	\caption{The logarithm of the production rates and A($\theta$=0$^\circ$)f$\rho$ ratios with respect to CN and to OH  as a function of the logarithm of the heliocentric distance of comet 41P. Pre-perihelion values are represented with filled symbols and post-perihelion values with open symbols. The log(Af$\rho$/Q) ratio is expressed in cm s/molecules.}
	\label{ratios_rates}
\end{figure*}

The logarithm of the production rates ratios with respect to CN and to OH as well as ratios of A(0)f$\rho$-to-Q(OH) and A(0)f$\rho$-to-Q(CN) for comet 41P are shown in Figure \ref{ratios_rates}. Generally, the ratios do not vary significantly on both sides of perihelion, but they vary significantly with respect to OH than with respect to CN. The C$_2$-to-CN production rate ratio (Figure \ref{ratios_rates}(d)) stays nearly constant with perihelion distance.

Determining a mean value for the dust-to-gas ratio is problematic for several reasons such as the dependence of Af$\rho$ on aperture size, heliocentric distance and phase angle. In this work, we have computed the Af$\rho$ values in the coma at 5000 km from the nucleus and we corrected it for the phase angle effect at $\theta$=0$^\circ$. The mean values for log[A(0)f$\rho$/Q(CN)] and log[A(0)f$\rho$/Q(OH)] are -23.00$\pm$0.06 and -25.46$\pm$0.03 cm s/molecules, respectively for comet 41P, and -22.88$\pm$0.12 and -25.66$\pm$0.11 for comet 45P.  As shown in Figure \ref{ratios_rates}(e) and (j), the evolution of the dust-to-gas ratios of 41P is symmetric on both sides of perihelion. The relatively low dust-to-gas ratio for 41P and 45P is consistent with the trend of an increasing dust-to-gas ratio as a function of the perihelion distance found by \cite{A'Hearn1995} (see Figure 4), which they assuming to be associated with thermal processing of nuclei surfaces.

Figure \ref{Ahearn} represents the logarithm of the ratio of the C$_2$ to OH production rates against the logarithm of the ratio of CN to OH, using a mean value for 41P  and 45P compared to JFCs and Long Period Comets observed with TS telescope between 2010 and 2016 \citep{Opitom2016thesis}, and also compared to typical and depleted comets given in \cite{A'Hearn1995}. Most comets observed with TRAPPIST, including 41P and 45P, are typical and lie on a diagonal trend, except for comet 168P/Hergenrother which is depleted in carbon chain species.

\section{Coma morphology} 
\label{sec4}

\begin{figure*}
	\centering
	\includegraphics[scale=0.55]{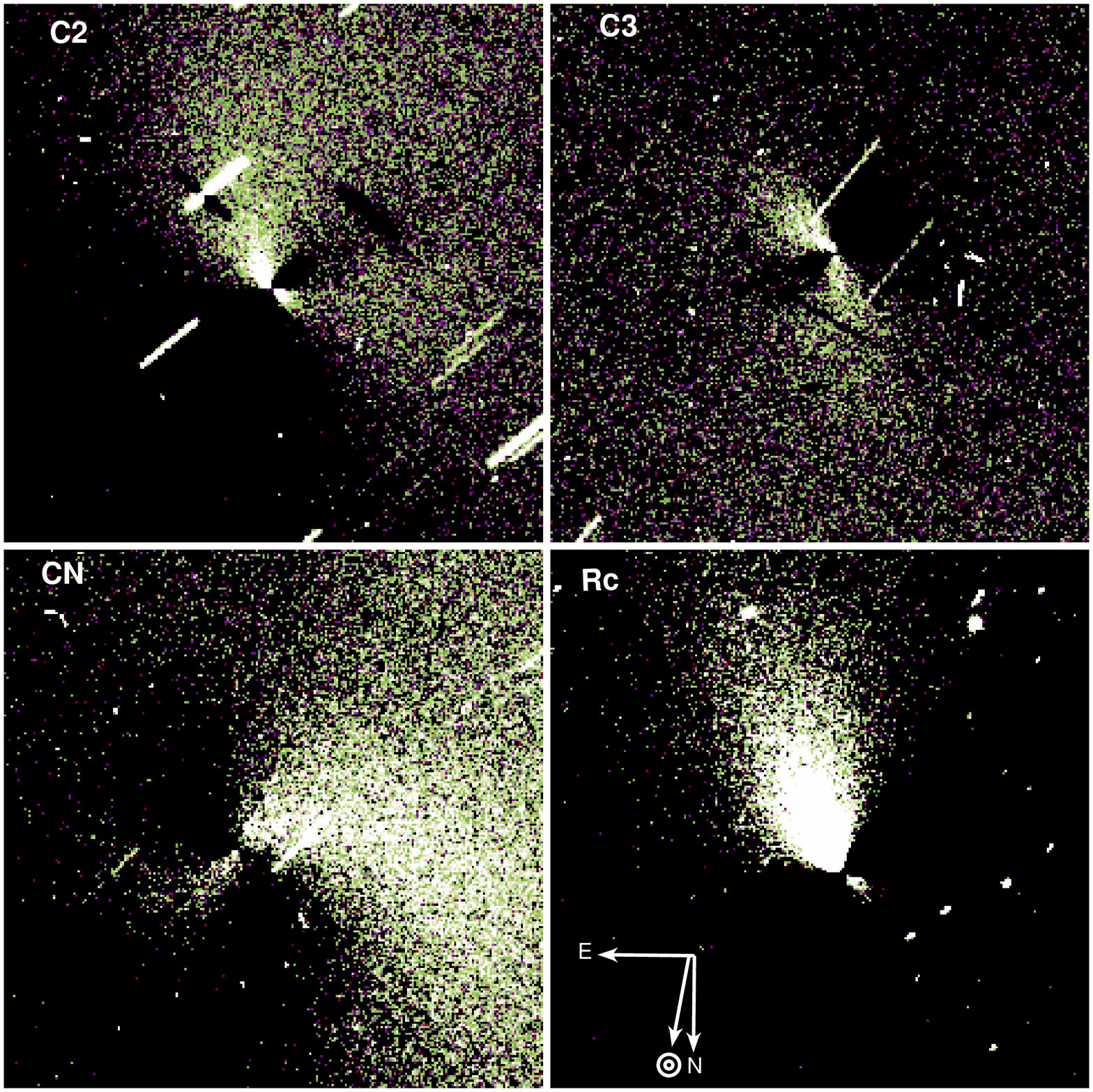}
	\caption{Gas and dust coma morphology of 41P on March 31, 2017. Each image is centered on the nucleus and enhanced by a simple rotational filter. The filter name is indicated at the top of each frame. All images are oriented with North down and East left with the direction of the Sun as indicated in the Rc frame. The field of view of each image is 5.8$^\prime \times$ 5.9$^\prime$, which covers 36600 $\times$ 37800 km.}
	\label{morphology}
\end{figure*}

In this section, we will discuss the morphology of the coma of 41P, which showed an important activity on both sides of perihelion. 45P does not show any interesting jets activity because it was far away from the Sun, when we started to monitor it.
 
The coma is dominated by the overall brightness distribution, and many individual features are not easily recognizable. This makes analyzing enhanced cometary images difficult, as important structural information is contained both near the coma center and in the dimmer outer regions. The study of coma morphology can give us information about the rotation period, active areas and homogeneity of the nucleus. This section is dedicated to the morphological features of gaseous species and broad-band Rc continuum images, using a simple rotational filter which takes the difference between two oppositely-rotated copies of the image. This technique requires an angle by which the image is rotated clockwise and counter-clockwise prior to calculating the difference. We tested several rotation angles, and they always revealed the same features. Finally we adopted a 40$^\circ$ angle, which provided the highest contrast images. The technique presented here is very sensitive to the centering of the nucleus, which we can set as the centroid of the profile or put manually.


Figure \ref{morphology} shows an example of CN, C$_3$, C$_2$, and Rc features on March 31, 2017. We decided to use the broad-band dust filter (Rc), even it is a bit contaminated by the gas, to show the dust features in our data because the SNR is too small in the narrow-band dust filters images for such processing. These images were obtained one week before perihelion. Usually, we acquired one or two C$_2$ and C$_3$ images per night, while we acquired more than 6 images for CN. Over most of the nights we saw CN and C$_2$ with sufficient signal-to-noise to detect variations of the jets positions in the coma caused by a variation of the viewing geometry. We started to detect the CN features at the beginning of March but we only saw one main jet, due to the low signal-to-noise of the images at that time. At the end of March, and as shown in Figure \ref{morphology}, we were able to see a second jet more clearly with the relative intensity and position of the jets varying with the rotation. This allowed us to measure the rotation period of 41P's nucleus (see section \ref{rotation_41P}). It is more or less the same feature that we saw in April with also a variation of jets position and brightness over the night. The C$_3$ and C$_2$ features are weaker than CN, they both display two short jets in opposite directions with different brightness compared to CN. They are also weaker than the dust jet (frame Rc), and are not detected in every image during our observations. The OH and NH features are not detected in our data set, due to the lower signal-to-noise ratio in these bands. The Rc image shows an enhancement of the coma in the dust tail direction.

\begin{figure*}[!]
	\centering
	\includegraphics[scale=0.585]{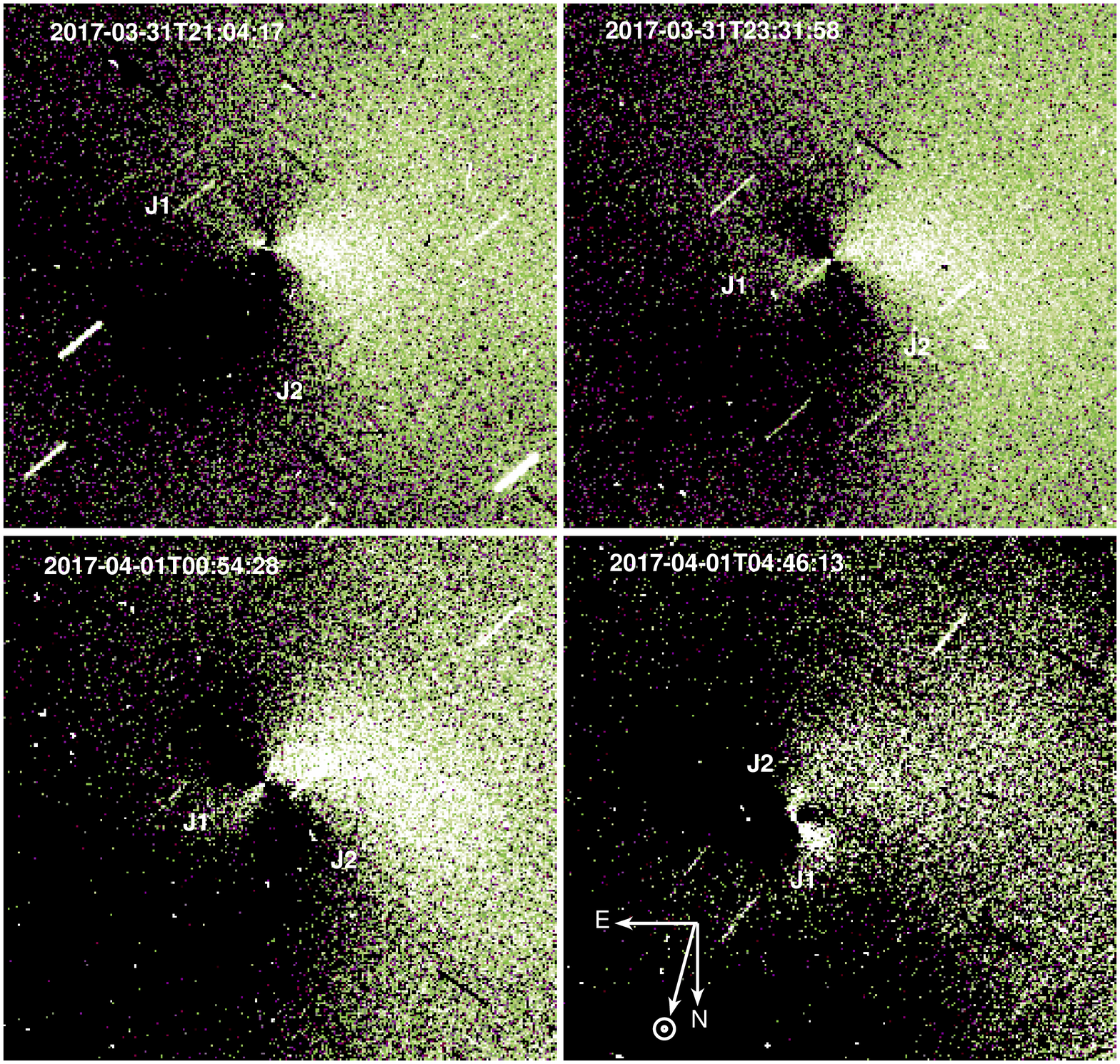}
	\caption{CN coma morphology of 41P on March 31, 2017 (before perihelion). Each image is centered on the nucleus, and has been enhanced by a simple rotational filter. The date (in UT) is indicated at the top of each frame. In all images North is down and East to the left with the direction of the Sun  indicated in the  bottom right frame. The field of view of each image is 8.5$^\prime \times$ 9.8$^\prime$, which covers 53120 $\times$ 26520 km.}
	\label{march_31}
\end{figure*}

\begin{figure*}[!]
	\centering
	\includegraphics[scale=0.455]{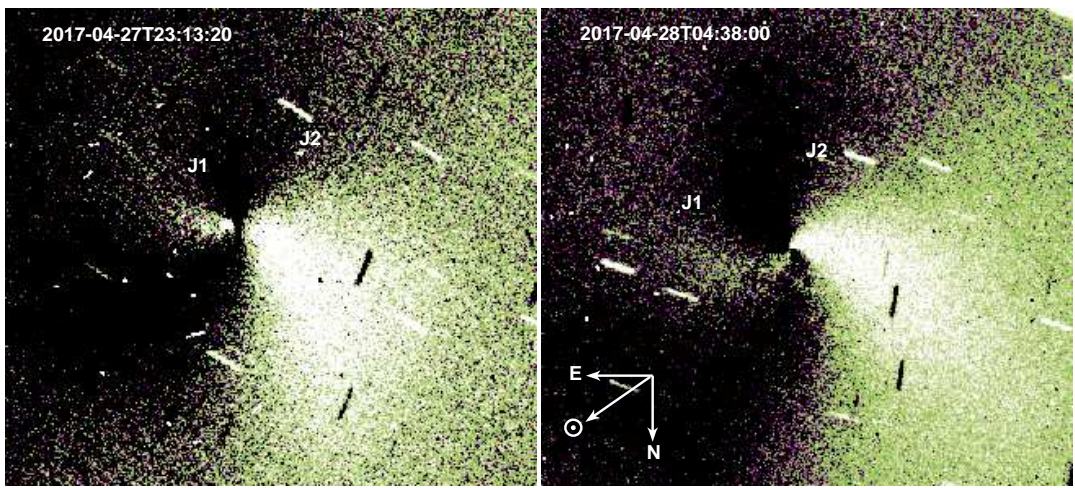}
	\caption{CN coma morphology of 41P on April 27, 2017 (after perihelion). The images are enhanced by a simple rotational filter. The date (in UT) is indicated at the top of each frame. In all images North is down and East to the left with the direction of the Sun as indicated in the right frame.  The field of view of each image is 8.4$^\prime \times$ 7.6$^\prime$, which covers 67500 $\times$ 62800 km.}
	\label{april_28}
\end{figure*}

\subsection{41P Rotation period}
\label{rotation_41P}

\begin{figure}[!]
	\includegraphics[scale=0.37]{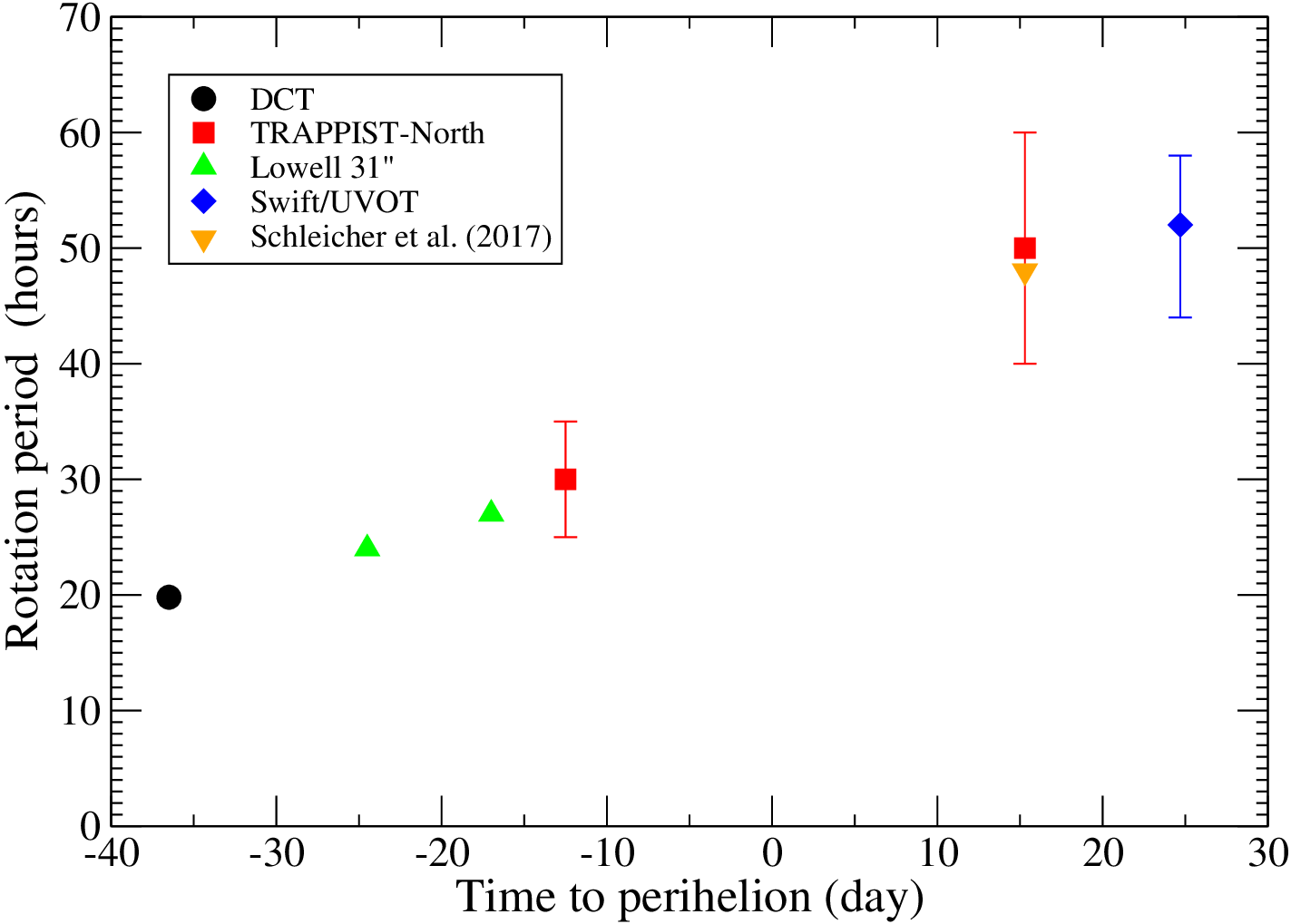}
	\caption{Rotation period evolution of comet 41P as a function of time to perihelion. The DCT value (black circle) is from \cite{Farnham2017}. The Lowell values (green triangles) are from \cite{Knight2017}. The triangle down value is from \cite{Schleicher2017}. The Swift value (blue diamond) is from \cite{Bodewits2018}. The error bar on the Swift observation  indicates the range of possible solutions due to the uncertainty in the change of activity as a function of heliocentric distance.}
	\label{rotation_period}
\end{figure}

The changes of orientation and morphology of the features is not necessarily a direct representation of the rotation period of the nucleus. The coma morphology depends on the shape, activity and rotational state of the nucleus of the comet, but can also be influenced by the orientation of the active regions on the nucleus, changing viewing geometry and seasonal changes in activity \citep{Belton1991,Jewitt1997,Davidsson2001,Keller2015}. However, observing repeating changes in the morphology of the coma is still one of the best way to determine the rotation period of a comet nucleus from ground based observations \citep{Samarasinha2004,Samarasinha2014}.

We enhanced the CN narrow-band images of 41P using a simple rotational filter, as described in the previous section. Often, CN images are used to investigate the gas coma features because those images have a better SNR and are less contaminated by the underlying dust continuum. This technique has been used to measure the rotation period of several comet nuclei such as 1P/Halley \citep{Ahearn1986}, C/2004 Q2 Machholz \citep{Farnham2007}, 8P/Tuttle \citep{Waniak2009} and C/2012 F6 Lemmon which was observed with TS \citep{Opitom2015a}. 

The first two series of CN images were obtained during the nights of the 3 and 7 of March, 2017, but it is hard to distinguish the jets in the coma because of the faint signal. After that, we do not have long series anymore until March 31, when we took a series of 16 images with exposure time of 600 s spanning over 8 hrs. We enhanced the images using the method described above. Figure \ref{march_31} shows an example of the CN coma features obtained. Two jets labeled J1 and J2 are clearly detected like partial spirals in a counter-clockwise rotation. The jets are rotating, with a position angle difference of 90$^\circ$ between the first and the last image of the series. Dividing the measured position angle difference by the time between the measurements, and considering that the jet is moving at the same rate throughout the entire rotation, we obtain a rotation period of about (30$\pm$5) hrs. The last long series is on April 27. We collected 25 images in the CN filter over 5 hrs. Figure \ref{april_28} shows two examples of CN enhanced images for this date. Two jets are detected (J2 brighter than J1) moving slowly in a counter-clockwise rotation. The change of the jets position angle is 40$^\circ$ over 5 hrs which gives a rotation period of (50$\pm$10) hrs. In conclusion, with this two long series that we obtained during one month, we found a change in the rotation period of about 20 hrs with an average increase of 0.68 hr per day.  

Several authors reported on the rotation period of the nucleus of 41P. \cite{Farnham2017} reported a rotation period of 19.75-20.05 hrs on the 6-9 March using images of CN coma  features  taken with the Large Monolithic Imager on Lowell  Observatory's 4.3-m Discovery Channel Telescope (DCT). Using the Lowell Observatory's 31'' telescope and the same method, \cite{Knight2017} reported change of the rotation period from 24 to 27 hrs during the 19-27 March period. Using the aperture photometry technique, \cite{Bodewits2018} found that the rotation period was changing from 46 to 60 hrs, with an average increase of 0.40-0.67 hr per day, during the period of 7-9 May using Swift/UVOT data \citep{Swift2004}. \cite{Schleicher2017} measured a rotation period of 48 hrs approximately on April 28. In gross, the rotation period of 41P more than doubled between March and May 2017, increasing from 20 to 50 hrs. Our measurements are in agreement with these data. Figure \ref{rotation_period} shows the evolution of the rotation period of 41P's nucleus as a function of time to perihelion. \cite{Bodewits2018} extrapolated the rotation period of the comet in time to investigate the past and future behavior assuming that the activity level and  torques did not change strongly during past perihelion passages, as well as the water production rates and orientation of spin axis. The authors found that before 2006 the comet could have been rotating with a period of only about 5 hrs, which is near the fragmentation limit. In the future they assume that the rotation period could exceed 100 hrs. They hypothesize that the rapid rotation might be linked to the bright outburst that occurred during the perihelion passage in 2001. This is the fastest rate of change ever measured for a comet nucleus. \cite{Pozuelos2018} found a complex ejection pattern which started as full isotropic, then switched to anisotropic about 1.21-1.11 au inbound (February 24 - March 14, 2017), and switched again to full isotropic around 1.3-1.45 au outbound. During the anisotropic ejection, they found that two strong active areas took over the dust emission, ejecting at least 90\% of particles. they related this result with the spin down found by \cite{Bodewits2018}, in the sense that those active areas could act as brakes. All of this may suggest the nucleus rapidly changes its rotation period as a result of changes in cometary activity and the strong active areas.

Several JFCs have also shown a variation of their rotation periods. The best example is 103P/Hartley 2, the target of NASA's EPOXI mission, which has a nucleus of 0.57 km and a rotation period of 17 hrs \citep{AHearn2011}. This comet showed a variation of its rotation period of 2 hrs in the three months around perihelion (October 28, 2010) during its return in 2010 \citep{Knight2011,Belton2013,Samarasinha2013,AHearn2011}. However, 103P had a peak water production rate three times higher than 41P \citep{Jehin2010,Knight2011}. In this context, we can also mention 9P/Tempel 1 (P$_{rot}$= 41 hrs) rotation decreasing by 0.2 hr over a period of 50 days \citep{A'Hearn1995,Belton2011,Samarasinha2013,Manfroid2007}. The target of ESA's Rosetta spacecraft, 67P (P$_{rot}$= 12 hrs) shows a decrease of $\sim$0.4 hr over 10 months after its perihelion passage \citep{Mottola2014,Keller2015}. Non-periodic comets have also shown changes in their rotation such as C/2001 K5 (LINEAR) \citep{Drahus2006} and C/1990 K1 (Levy) \citep{Schleicher1991,Feldman1992}. Many have shown a rotation period change, even though the amplitude in the case of 41P is rather unusual.

\section{Summary}
\label{sec5}

We have presented the results of the photometric monitoring   and imaging of the JFCs, 41P and 45P, over a five months period with TRAPPIST-North telescope. OH, NH, CN, C$_2$ and C$_3$ production rates were computed as well as A($\theta$)f$\rho$ parameter on both sides of perihelion for comet 41P and after perihelion for comet 45P. 41P and 45P are dust poor comets in comparison with others JFCs. They have a "typical" composition regarding their Q(C$_2$)/Q(CN) and Q(C$_3$)/Q(CN) ratios. We have shown that the activity of 41P is decreasing by about 30\% to 40\% from one apparition to the next. We analyzed 41P coma features in CN, C$_2$ and C$_3$ filters as well as those of the dust continuum windows using a simple rotational filter. We detected variations of the coma morphology during long series of observations performed over individual nights, due to the rotation of the nucleus. This allowed us to determine the rotation period of 41P, which increased from (30$\pm$5) hrs at the end of March to (50$\pm$10) hrs at the end of April, when the comet was closer to the Sun. Our results confirm the strong braking of 41P rotation presented in the literature \citep{Bodewits2018,Knight2017}.

\section*{Acknowledgments}

TRAPPIST-North is a project funded by the University of Liège, in collaboration with Cadi Ayyad University of Marrakech (Morocco). TRAPPIST-South is a project funded by the Belgian Fonds (National) de la Recherche Scientifique (F.R.S.-FNRS) under grant FRFC 2.5.594.09.F, with the participation of the Swiss National Science Foundation(FNS/SNSF). Y. Moulane acknowledges the support of Erasmus+ International Credit Mobility. E. Jehin is FNRS Senior Research Associates. F.J. Pozuelos is a Marie Curie Cofund fellow, cofunded by European Union and University of Liège. M. Gillon is FNRS Research Associate and Jean Manfroid is Honorary Research Director of the FNRS. We thanks NASA, David Schleicher and the Lowell Observatory for the loan of a set of HB comet filters.

		
\bibliographystyle{aa}
\bibliography{biblio.bib}
\end{document}